# An Analysis of Impact Pathways arising from a Mobile-based Community Media Platform in Rural India

**Aparna Moitra**
University of Toronto
Toronto, Canada
aparna.moitra@utoronto.ca

**Archna Kumar**
University of Delhi
New Delhi, India
archnak01@gmail.com

**Aaditeshwar Seth**
Gram Vaani and IIT Delhi
New Delhi, India
aseth@gramvaani.org

## ABSTRACT
Our research presents the case-study of a mobile phone based, voice-driven platform – Mobile Vaani (MV), established with a goal to empower poor and marginalized communities to create their own local media. In this paper, we derive a comprehensive theory of change for MV from the data gathered using the Most Significant Change technique. This paper contributes towards formulating a theory of change for technology-driven community media platforms which can be adapted to other ICTD interventions too.

## CCS Concepts
• K.4: Computers and Society~ K.4.3: Organizational Impact

## Keywords
community media; IVR; mobile phones; social impact

## 1. INTRODUCTION
Our research presents the case-study of a mobile phone based, voice-driven platform – Mobile Vaani (MV), established by the social enterprise Gram Vaani (GV) with a goal to empower poor and marginalized communities to create their own local media [31,32]. Technologically, MV operates using an IVR system which avoids the need for users to have a data connection or smartphones. Users can simply place a missed-call to the MV phone number, and the IVR server disconnects the call and calls them back. The platform therefore remains free-of-cost for the users. Using phone key-presses for navigation, users can then record voice messages they want to share, and listen to messages left by others, making the platform suitable for use as an interactive discussion forum even among low-literacy users. MV started operating at a small scale in 2011 and since then has grown to cover more than 25 districts across the three states of Bihar, Jharkhand, and Madhya Pradesh in India. MV has serviced over 10,000 calls per day from 100,000 unique users monthly at its peak, before it had to scale down a bit for funding limitations. On a daily basis, users contribute 400–500 messages on topics such as local news, agriculture, health, career counseling, job postings, discussions on gender empowerment and social norms, cultural events and folk songs. All content on the platform is reviewed by a team of content moderators, who listen to all user contributions and determine whether to publish, reject, or edit the contributions.

India's states of Bihar, Jharkhand, and Madhya Pradesh, which comprise MV's main areas of operation, are among the most poverty-stricken and low-literate states [34]. Citizens there also suffer from inequality ingrained through sociocultural norms such as patriarchy and caste structures, political dynamics, large-scale corruption, mis-governance, and left-wing extremism [10,32]. The population is predominantly rural and remote, lacking easy access to conventional media platforms such as TV, radio, and newspapers but has significant access to mobile phones due to their deepening penetration in the region [31]. Therefore, MV attempts to offer new communication spaces easily accessible to these populations. MV socially embeds itself within the communities it serves by engaging local community members as volunteers to mobilize new users. These volunteers train and encourage new users to use the technology and also develop appropriate use-cases for the users to regularly use the MV platform [32].

MV also performs functions that go beyond just serving the local community on handling local issues. For instance, it attempts to build public opinion on current affairs and public policy by soliciting topics on current affairs, national policies, local events, etc. from the users (with the help of volunteers) and initiates a discourse [41] in the community around these topics via its regularly scheduled programmes. The participatory and bottom-up (two-way and multi-way) nature of the platform highlight users' diverse viewpoints.

Our objective from this paper is to formulate a theory of change (ToC) which explains the processes behind the learnings and agency effects reported by MV users upon their engagement with the platform. For this study, we collected data over a five-year period from 2011-2016 and followed a qualitative methodology which involved the use of the Most Significant Change technique [9] and field immersion elaborated in the methods section. Upon analyzing our data three distinct elements of study emerged: a) platform processes; b) platform characteristics; and c) learning and agency effects.

MV's *platform processes,* comprised of methods for content generation, community mobilization and building institutional linkages, can be considered as key programmatic inputs to bring about social change. These inputs operated under four key *platform characteristics* or norms, (i) content relevance; (ii) editorial credibility; (iii) dialoguing opportunities; and (iv) multi-level engagement processes, that were found to be essential to bring about social change. These characteristics were also found to be essential to lead to the effects or changes that users in their stories described as the changes they felt from their engagement with the platform. We finally categorized these effects as different kinds of *learning* and *agency effects* experienced by the users, and use a range of theories on technology and development to explain our observations. We finally use the concept of Principal Communicative Actions [30,18] to stitch the ToC as a sequence of platform processes acting as inputs, platform characteristics acting as outputs and the effects acting as outcomes for long-term change.

We begin with a discussion of related work in the next section, followed by a description of the research methodology in detail. We then present our findings and finally outline the theory of change emerging from the paper. We also discuss how MV's ToC can be generalized and applied by other ICTD initiatives to bring about change among individuals and communities.

## 2. RELATED WORK
Impact assessment of initiatives forms a key area of discussion within the ICTD community with several researches having outlined techniques to assess impact [15,19,24,27,28]. Gigler [15] and Kleine [24] have treated ICTs in general as an artifact and use combinations of the Capability approach [38] and Sustainable Livelihoods framework [2] to theorize how ICTs expand individual and collective capabilities of their users.

In the existing literature, impact measurement has been carried out in both quantitative and qualitative way. Quantitative methods include baseline-endline surveys and RCTs (Randomized Control Trials). Studies in the qualitative domain have used a wide array of approaches such as interviews, focus groups, content analysis, case study, storytelling and ethnography to measure and demonstrate impact in more qualitative terms. More specifically, to evaluate the impact of IVR based systems in ICTD, Marathe et al. [27] have attempted to document the impact of CG Net Swara, an IVR based grievance redressal platform active in rural parts of central India, and have compared its impact with government run grievance helplines [28]. However, they stopped short of outlining a theory of change for the impact. Our paper attempts to fill this gap.

The pathways outlined in the MV ToC are based on the principles of successful online and offline strategies employed by various ICT- and non-ICT-driven projects to bring about social change. First, the pathways borrow insights from the studies of technology adoption which have shown that users will only adopt technology if their social environment encourages its uptake (e.g. encouragement by friends and family to access entertainment, relevant information, etc.) [25,43] and will use it sustainably if these technologies constantly adapt to their needs [31] or if they are constantly able to appropriate these technologies to suit their needs [32,26]. Second, contextual user generated content broadcasted / telecasted on the different media platforms coupled with offline mobilization techniques [13,48,46] increases the chances of users understanding that content, internalizing it and applying it to their circumstances. Third, similar to the news dissemination and opinion formation function of mass media platforms [21], technology platforms are able to disseminate information, local news, and hold discussions [17,31,33] that influence people's opinions on various current affairs and public policy matters. Fourth, technology platforms are able to amplify their users' issues and build pressure on relevant stakeholders via the platform (similar to the watchdog role played by mass media in a democracy). Localized tech platforms additionally form the epicenters to initiate on-ground processes of collective action among their users and are capable of building formal / informal linkages with various stakeholder groups, to perform a social accountability function [6]. All these pathways eventually converge on the idea that technologies amplify human intent [45] which individuals can use to bring about changes in themselves and within their communities. Therefore, building over these already existing ICT, non-ICT, online and offline methods, in this paper, we compile a list of pathways qualifying our ToC.

MV is a participatory media platform. Various scholars and activists [8,3,5] argue that ICT initiatives for participatory media have the potential to enable a digital version of Habermasian Public Sphere which creates ideal speech conditions for its users, enabling them to define their own agenda and openly participate in discussions, thereby leading to communicative action [18,23]. Users participation on such ICTD platforms thus influences their perceptions, helps them understand alternative viewpoints, arrive at a consensus and wherever required, act against injustice, forming the basis for change and empowerment [30]. In our work we demonstrate how MV's platform processes and characteristics help construe it as a digital public sphere thereby initiating these pathways for change.

## 3. RESEARCH METHODOLOGY

Our research objective is to formulate a theory of change for MV as a participatory media platform to bring about change among individuals and communities. We employed the qualitative methods of field immersion and the Most Significant Change Technique to understand MV's processes, users interaction with these processes and if the users' perceived any changes within themselves and among their community members due to their engagement with the platform.

Data was collected over a five-year period from 2011 until 2016. Throughout this period two of the authors were engaged in field immersion – one leading on all operational fronts and another focusing on MV's content processes with some time spent on studying MV's community mobilization processes. To ensure objectivity, yet another author was engaged to provide an outsider's perspective.

From 2013 onwards, we began enquiring if the users felt any changes upon engaging with the platform. For doing this we used a participatory, narrative-based technique known as the Most Significant Change (MSC) Technique [9], which enables individual community members to share their experiences of association with the platform and the changes in their lives they realized from their participation. In this method, stories are read aloud during focused group discussions among the community members and key stakeholder groups. The participants then deliberate on the outcomes mentioned in the stories to systematically voice their ideas and develop consensus about the outcomes they most value.

We collected a total of 98 significant change stories from MV users who were men and women aged 15–60. Among them, the majority were affiliated with lower-caste and tribal categories and came from an economically poor background. These stories were collected using an event-based sampling technique wherein the first author of the paper began engaging with those MV users and volunteers who showed up at MV's monthly district-level meetings during the bi-annual field visits. [51] provides in-depth details about the methods. While some of the users chose to write their stories of association with MV on their own, others chose to audio record their accounts in their local dialects of Hindi and Bengali. All these stories were transcribed and translated to English.

To provide assistance with the coding process for data analysis, we used a qualitative data analysis software - Atlas.ti and employed an inductive approach [49] to iteratively code and analyze the data for patterns, over three rounds. In the first round we carefully read and subjected the data to open coding. Upon its completion, we identified three broad themes of – i) platform processes; ii) platform characteristics; and iii) effects of association emerging from users' narratives. The themes were assigned to the data and were color coded for convenient identification. In the second round, we further explored the data for sub-themes and axially coded them to the main themes. An exhaustive list of 125 codes was identified (e.g. skill building, value of user generated content, improved awareness about child marriage, etc.) and a codebook was developed by the first two authors. All the authors presided to code the data together as a group which allowed us to share our reasons for utilizing the codes in particular ways, share examples and non-examples of a code and reach consensus on the coding procedure protocol [50]. In the third round, upon many iterations of inductive analysis, we were finally able to condense our coded data into a schema, that we present in detail in the findings section of the paper.

We use an assortment of theories, concepts and frameworks of Participatory Communication, Empowerment, and Development [15, 30, 40] and theories of community media [3] to further analyze and interpret our findings to situate them within the broader body of work. We use the theory of Principal Communicative Action for outlining the theory of change because of MV's alignment with the idea of Habermasian public sphere, ideal speech conditions and its potential to bring about change.

## 4. FINDINGS

We present our findings in three parts. First, we briefly explain MV's core operational processes – also known as the platform processes – of content generation, community mobilization and building institutional linkages in Section 4.1. We then elaborate on platform characteristics that emerge from these platform processes in Section 4.2, which according to the users motivate them to participate on MV. Finally, we outline the effects of users' participation on the platform in Section 4.3 where they are classified as different kinds of individual learning and collective agency effects.

## 4.1 Platform Processes

MV is driven by three key processes of content generation, community mobilization and building institutional linkages.

First, the content generation processes facilitate the management of User Generated Content (UGC) which is audio recorded on MV and is reviewed by a team of content moderators for editorial sanity. The UGC is usually hyper-local in nature carrying local news, information on local events, folk expressions, etc. The content processes also ensure that the content on the platform is as per users' needs by producing and curating some content in-house on the topics of agriculture, health, career counseling, job postings, etc. The in-house produced content also initiates discussions on the issues of gender empowerment and social norms as well as on the largely *glocal* [11] issues of public policy.

Second, the process of community mobilization involves identification and training of volunteers from the local communities to mobilize new users. These new users need to be trained on using the platform as well as need to be encouraged and coaxed to become habituated to use the platform. The community mobilization processes hence drive technology adoption, hyper-local content generation, local reporting, and build local institutional linkages to help in users' grievance resolution.

Finally, the process of building institutional linkages among various stakeholders at different levels not only helps in advocating for and resolving the problems users face in accessing government schemes and public utilities but also helps in providing access to expert information on various issues. These processes are explained in our previous work [31] and we do not detail them in this paper. Here, we focus on users' interaction with these processes and their role in deriving the ToC for MV.

## 4.2 Platform Characteristics

Based on the users' responses, we were able to identify four *platform characteristics* of – (i) content relevance; (ii) editorial credibility; (iii) dialoguing opportunities; and (iv) multi-level engagement processes that have emerged from the *platform processes* described above, and influence users' participation on the platform. We now describe each of these platform characteristics in detail.

**Figure 1: The Mobile Vaani Theory of Change (ToC)**

### 4.2.1 Content Relevance and Relatability

Many users in their stories acknowledged that MV carries localized content based on their topics of interest such as local news, event updates, entertainment and cultural content, local government schemes, information on health, agriculture, local employment opportunities, as well as self-composed, folk poetry and songs, programmes for children etc., which they find more relevant than content from other media platforms available to them. 53-year-old MV user from Baghmara, Dhanbad in Jharkhand elaborates these points in his story.

*"...MV provides me with content about my region that I don't get out of watching TV. I get all the information about the local government schemes, their implementation, education, health, employment related information, local event updates and entertainment and cultural content from MV. We watch programs on similar issues on TV, but Jharkhand related content is very less... and since they are not from our local scenario, we don't find them very relevant. The newspapers are also mostly filled with the news from town and district centers, state and national issues. The coverage of our village level issues is very limited..."*

It is important to note that when people have such an opportunity to create and share content on the platform in their own language it helps in contextualizing even the most difficult concepts and makes that content more relatable and actionable for the fellow listeners. Many users reported that listening to UGC and local news on MV helped them better understand current affairs and how national policy level issues affected them. Story analysis also brought out that the presence of local cultural symbols and entertainment-education (EE) format [42] in MV's content especially appealed to the users. While these cultural symbols were inherent to the UGC such as folk songs, poetry and narratives; MV's content team also situated the content they curated and produced within the users' local context by using commentary, interview, and radio dramas in the EE format to portray personal challenges and family situations prevalent within the user community. Users in their stories reported that while this helped them contextualize complex social issues such as gender-based violence, superstition, etc. making it more relatable to them, the use of the EE format made their experience of listening to MV more pleasant, enjoyable and meaningful.

### 4.2.2 Editorial credibility

Credibility of the content published on MV formed another key characteristic of the platform. An analysis of users' stories revealed that they trusted the authenticity of the content on the platform for two reasons – (i) MV's editorial processes; and (ii) the presence of diverse stakeholder groups that contribute wide-ranging views to promote information completeness on the platform.

MV has editorial processes in place to consistently check for the credibility of content recorded on the platform. These processes of the platform to practice minimal editing and reject only those content items that are found to be objectionable in nature (e.g. use of abusive language; unverified, religiously and politically motivated content promoting religious and ideological bias etc.) were appreciated by the users. Users were of the view that authenticity of their content was maintained on MV and that the content they recorded was not edited to suit any one idea or edited to such a degree that its meaning was misconstrued. In his story, 30-year-old MV user from Topchanchi, Dhanbad, in Jharkhand commented upon the editorial freedoms that MV provided him with to voice his ideas without editing or inappropriately using his comments out of context.

*"MV provides us with an independent stage to discuss what we want to, demand accountability, initiate a dialogue and voice our opinions on topics of our choice and in our own words. It also gives us the space and a decent amount of airtime to share our views with*

*others… the entire item (content) goes through word-by-word just as we record it…"*

In their stories, many users particularly reported valuing the airtime they received on MV to express themselves on topics of their choice, in their own words, which the mainstream media platforms could not provide them. In fact, some users who had the experience of engaging with the stringers of media houses at the grassroots level shared that it was immensely difficult to get their issues covered over the mainstream media platforms due to the gatekeeping processes. This is understandable due to an immense volume of news being generated. However, a few MV users also reported that it was common for stringers to coerce them into paying bribes to get their news published as the stringers were reluctant to carry news on serious matters that were not sensational in nature.

In this context, users appreciated the editorial openness of MV that promoted their freedom of expression, as opposed to mainstream media's inability to adequately represent all the issues and diverse perspectives. As an example, several stories reflected that users appreciated the unconstrained and triangulated representation provided to para-teachers' (government school teachers on contract) strike on MV which received inadequate coverage on most mass media platforms. By means of publishing UGC from para-teachers, students, and their parents/guardians expressing their side of the story, MV was able to present the issue from the perspectives of all stakeholders conducting and affected by this strike [31].

Thus, MV's editorial processes ensure greater context and completeness by publishing UGC [41]. The presence of multiple stakeholders on the platform provided diverse viewpoints and expertise on the issue to increase the completeness of its coverage. This diversity of perspectives was made more explicit with the help of content curation processes that sometimes deliberately placed diverse perspectives together, to have the users review information from diametrically opposite viewpoints.

### 4.2.3 Dialoguing Opportunities

Users in their stories reported that MV enabled an easy and more inclusive exchange of dialogue on the platform – by virtue of its technology as well as its editorial policies, which we detail here.

First, since the moderation happens round the clock in shifts distributed across different moderators, MV enables rapid publishing of UGC as they are recorded. This enables users to enter into an impromptu "non face-to-face" dialogue which is self (user)-paced in nature. Many users reported that they felt the speed of content publication on MV to be just as fast as that of news on TV and much faster than a newspaper. In their stories, users reported that the speedy publication of content from a multitude of districts across the state, made them aware of what was happening in their own surroundings as well as kept them appraised of local events unfurling in far-away districts. Thus, the virtual medium reduced space-time barriers as the users could access, comment and engage in a dialogue on any issue of choice at their convenience.

Second, the content and editorial mechanisms of the platform create multiple opportunities for engaging users in a dialogue on the platform. These mechanisms are able to – (a) identify trending topics on the platform and initiate directed discussions around them; (b) cross-post interesting UGC across clubs[1] of different districts to initiate discussions; (c) initiate informational and activism based campaigns on different social issues [31]; (d) initiate dialogue on fun and entertaining topics, e.g. trivia; and (e) solicit topics (with the help of volunteers) that users find difficult to understand, and initiate discourse [41] in the community around these topics via its regularly scheduled programmes such as *Janta ki Report* (People's report). This programme enables users to comment on each other's opinions and is explained later in more detail in the introspectional learning section of the paper.

These opportunities enable users to share their views as well as express dissent or affirmation of ideas by composing songs and poetry or; use satire to lightly share the content of a more serious nature or; use direct narrative-based content to communicate their issues and challenges. This leads to the emergence of a participatory dialogue which happens when individuals listen to each other, speak with each other and share the dialogue space with respect and consideration such that it leads to the creation of a fair and just space, and the inclusion of traditionally and historically excluded stakeholders [12, 20].

Users who shared their stories acknowledged that MV reflected these dialoguing characteristics of being an inclusive and a fair and just platform. They found MV to be inclusive because it facilitated the participation of different stakeholder groups from all sections of society and gave them the space to articulate their ideas [31]. Their stories emphasized how MV's mobilization strategies especially focused on enabling the participation of vulnerable and marginalized groups. Similarly, users attested to MV being a fair and a just space as they can freely use the platform to share their opinions and experiences as well as call out injustice in the matters of denial of their rights and entitlements, discussed in detail later, without having to worry about a pushback from powerful community members.

Hemmati [20] refers to such multi-stakeholder dialoguing process as being *generative* in nature because diverse views and ideas that enable exchange in a respectful and creative way lead to a complete understanding of actions and solutions that wouldn't have been possible in any other way. These dialoguing characteristics also fulfilled Habermas's [18] ideal speech conditions of – (a) symmetric distribution of opportunities to contribute; (b) freedom to raise any proposition and; (c) equal treatment of propositions raised, deemed essential for participatory dialogue [23].

However, we have also noted a couple of exceptions to this inclusive, fair and just characteristic of the platform [32]. First, given the patriarchal context of MV's areas of operation, the platform is mostly male-dominated, and it has been a challenge to include women in the dialoguing processes. While the institution of women-only volunteer clubs has helped in the creation of separate spaces for women, but socio-cultural barriers that limit women's access to phones, technology and public spaces in general have prevented the mobilization of an equivalent proportion of women. Second, there have been instances where users and volunteers belonging to dominant caste groups have objected to the inclusion of lower caste users and volunteers on the platform on the grounds of them being unable to speak clearly or submit quality content. While the club structure has managed to control this to a significant extent, these social prejudices remain in the backdrop as the unresolvable knots deterring change.

---

[1] GV split the state-level MV platform into district wise groups, known as clubs. Each club has a different name and phone number indicating that it is a local wing of the statewide MV.

### *4.2.4 Multi-level engagement processes*

MV offers three escalating levels of user engagement – platform mediated; volunteer mediated and; institutionally mediated engagement – which seemed to influence users' participation on the platform the most. These processes enabled users to develop their own use-cases of using the platform, as per their needs. This was especially relevant for the users because within their larger socio-cultural context, the systemic denial of their rights and entitlements left them with a ubiquitous need to seek accountability towards the resolution of their issues and challenges. They were hence expectant of the possibility of resolving their issues through the platform. By developing accountability seeking processes with their volunteers to engage with local stakeholder groups as well as establish linkages with a variety of institutions, MV extended the scope of the platform and its potential to address its users' issues and demand accountability from government departments and other concerned entities. These multi-level engagement processes were found to be especially valuable and helpful in embedding MV platform among its users. This further strengthened users' participation on the platform and built ownership and credibility of the platform among them. In this section, we discuss how the inherent capacity of the platform to amplify each user's concern was coupled with different organizational processes set up for them to seek recourse to their problems and demand their rights and entitlements.

- *Primary level: Platform mediated engagement* – At this level, user engagement is limited to the platform as the platform characteristically enables voice amplification at no cost to the user. By sharing their grievances on the platform, users find support and affirmation and the accountability processes at this level stem organically. Thus, MV's role at this level was only restricted to the users' engagement with the platform, with all other aspects emerging on their own. Several instances in stories portrayed users' efforts to advocate on issues such as – irregularity in the disbursement of entitled ration under the subsidized ration scheme; non-disbursement of wages of people working under the national rural employment guarantee scheme; corruption in the construction of local infrastructural amenities etc. – and rally against people in positions of power by reporting about their malpractices on the MV platform. This led them to garner support online on the platform as well as offline among their community members.

- *Secondary level: Volunteer mediated engagement* – To compliment primary level platform processes, hyper-local accountability processes were also initiated by the district level club volunteers in their local communities. In the beginning, volunteers would collectively forward individual grievances of people from their areas to the local administration and other concerned stakeholders. They then developed a local level network that lobbied for issues people faced locally. They worked hard to maintain a conducive relationship with these authorities so that their issues could be given due attention. In due course of time, this process was formalized to initiate accountability driven programmes such as *Janta Darbar* on their district-level club MVs where volunteers utilized their relationship with the local government officials to seek their commitment on issue resolution.

People's grievances and complaints were formally lodged with the concerned authorities during the *openhouses* which were held at the community level on a weekly basis.

At a more expansive level, the volunteer clubs initiated a monthly programme called *Jan Shakti Abhiyaan*, where more complex community issues raised by the users over the platform could be registered for petitioning through the IVR. Users were able to register their endorsement for the petition by pressing a key. The response on petitions strengthened the club volunteers to formally draft a letter on behalf of the club citing the count of signatories and transcriptions of some user experiences, and send it to the administrative authorities for action. Petitions on specific aspects of various government programs such as Mid-day meal scheme (MDMS), Public Distribution System (PDS), Mahatma Gandhi National Rural Employment Guarantee Scheme (NREGA), etc. have been floated thus far.

Many users in their stories appreciated these efforts by the volunteers, as 36-year-old MV user from Musabani, East Singhbhum in Jharkhand reflects in his story,

*"… I shared my report about stopping of mid-day meals at a nearby school since more than a month because of which the students stopped going to school. Afterwards, I got a call from Mobile Vaani volunteers asking about the details of my complaint. Using the details gathered from me and other users, they not only created a Jan Shakti Abhiyan petition on it, but also shared the interviews of concerned officials on this issue. I liked the accountability factor of this program and I feel that MV's efforts like these help us to openly question the system…"*

Thus, these volunteer mediated local level activities enabled a routine follow-up on these issues, which in turn built pressure on government functionaries and other stakeholders to address them.

Taking a cue from these accountability seeking processes, volunteers began to conceptualize similar linkages with experts and other local stakeholder groups to gather information about a range of issues concerning users in their community. For instance, information about local jobs to prevent migration, expert advice on farming to increase yield etc. were played on different club MVs by connecting with the appropriate stakeholders.

- *Tertiary level: Institutionally mediated engagement* – At this level, MV with the help of its parent organization GV, initiated linkages specifically with senior officials of government departments at the state level, so that they recognize issues reported on MV as legitimate grievances and work towards their resolution. Additionally, MV/GV also established partnerships with regional mainstream media organizations as well as civil society organizations to advocate for issues and further build pressure on those institutions with whom a linkage could not be worked out. Issues were escalated to this level when they were too complex to be resolved by the volunteer linkages.

Users in their stories acknowledged the efforts of MV's multi-level engagement processes to advocate issues at a macro-level for social causes and shifts in policy. This includes examples such as advocacy done with policymakers at the local level for migration due to unemployment; better quality of higher education in villages; rights of those who have been forcibly displaced due to mining and other development projects, etc. This provided people with the hope that their issues will be heard and resolved, and also acted as a great motivator to sustain their usage of the platform.

Thus, in order to drive users towards a sustained usage of the platform, MV went beyond its scope as a media platform to develop online and offline processes with different stakeholders and institutions to seek accountability and necessary action. This platform characteristic portrayed the democratic nature of the platform that made it easier for every user to seek accountability and vital information on their own. As the scope of addressing and resolving their issues and concerns increased with each escalating level, it further encouraged users to keep coming back to the platform. This aspect played a critical role in influencing people's participation on the platform as it resulted in the growth of the number of users on the platform, and also led the users to report greater ownership of the platform. At the same time, if MV was not able to work out a suitable linkage with the concerned authorites or stakeholders, it also led to several users being dissatisfied especially during the initial years of the platform when their expectations of the platform (to resolve their grievances) were not met. This was eventually countered by helping the users realize that MV was not a grievance redressal helpline and could not guarantee a resolution.

## 4.3 Effects of Association with MV

The ability of *platform characteristics* to continually engage people to participate on MV led its users to experience certain effects, both individually and collectively. These users, in their stories, identified different aspects of engaging with MV that they felt empowered them. We classify these effects as different kinds of (individual) *learning* and (collective) *agency effects*. In this section we present these effects arising from users' and volunteers' participation on the MV platform.

### *4.3.1 Learning*

Users' stories pointed towards a range of learning that emerged from the content they heard on the platform. According to their stories, these learning were central to their motivation to participate on the platform. We have classified these as – (i) informational learning; (ii) introspectional learning; (iii) instrumental learning and; (iv) social learning – with each type having several dimensions as well. We have operationalized many dimensions of these learnings using a range of references and theoretical frameworks which we will now elaborate.

*i. Informational Learning* – Due to the multiplicity of information shared on the platform, stemming from the platform characteristics of *content relevance and relatability* and *editorial credibility,* users found opportunities that enabled them to add fresh dimensions to their existing knowledge as well as learn about new issues and aspects that were previously unknown to them. They perceived information gain in diverse aspects of their socio-cultural, economic and political environment. Their usability of this information was further enriched by the platform characteristic of *dialoguing opportunities* that ensued on the platform as it enabled its interpretation in context of their lives and led to the generation of new perspectives and knowledge. The instances that reflect the various aspects of information learning due to users' association with MV are discussed as follows.

- Socio-cultural awareness - This aspect of users' learning focuses upon a deeper understanding of their cultural heritage such as local art forms, customs and traditions; superstitious and corrosive practices prevailing in their communities; their life-cycle practices evolving with time; as well as ecological awareness.

Many users who shared their stories reported an increased awareness of their cultural roots, as they came into contact with other users who were traditionally more aware of their folk culture (such as folk artistes - singers, poets, dancers; village elders etc.) via MV. Identifying different art forms and sharing information about them led to an enhanced interest in their popularity and retrieval amongst the community members.

Users also expressed that they were able to understand and learn more about their customs and traditions due to culture specific discussions initiated on MV on occasions of different local festivals, as well as during regular scheduled programs such as *Bole Dil ki Baat*. These discussions led different users to present what they know about the topic/issue, thus triangulating different perspectives about the genesis of their customs and traditions, and the reasons behind them. This in turn helped all the listeners understand these customs and traditions comprehensively, from different perspectives.

Another key aspect emerging from the stories was enhanced awareness and understanding about superstition and other practices related to various issues prevalent in the community. Users attributed to the discussions on MV their better understanding of the original logic and rationale behind these practices, and how over time these became superstitious and rigid. These discussions, for instance, explained why "caste-based bad omens" came into being – to establish the dominance of upper caste community members over those belonging to the lower caste to oppress them, and further promote their exclusion from the community. Likewise, these discussions rationalized the logic behind the superstitious practice of "witch-hunting" common in the villages of Jharkhand where women are branded as witches – raped, tortured, murdered and held responsible for illness, death or drought – by powerful members of the community to grab their land when they are the sole owner of a property, or to settle scores with them. Owing to the experiences shared within these discussions, many users in their stories appreciated MV's in-depth, and evidence based directed programming attempts at informing them about the actual logic behind these superstitious beliefs. Users, similarly, appreciated MV's regular informational campaign and programming focus on generating awareness about gender-based corrosive practices of dowry and child marriage for initiating a discussion on such widely prevalent but seldom discussed issues.

Users across different stages in their life-cycles reflected in their stories how their engagement with MV provided them with essential information that enabled them in better comprehending the issues of their own life-cycle stages. For instance, men in their roles of fathers valued parenting advice and understood the role of a father in children's upbringing; women in their roles as home makers and mothers valued the information on nutrition, health and parenting; and adolescents and youth valued information about reproductive and sexual health issues, personal hygiene and sanitation as well as issues they were negotiating such as relationships, marriage, exam stress and information about probable careers.

30-year-old MV user from Ghatshila, East Singhbhum, Jharkhand in her story shared how the campaign on parenting practices led to a change in her parenting approach.

*"…Slapping and physically beating children to discipline them is very common in our culture. I used to do the same until I heard its negative effects on children in a parenting campaign drama episode on MV. The situation was portrayed exactly how it happens between me and my child, so I was instantly able to relate to it. Since then I try my best not to punish my child physically and try to make him understand his misdeeds in a friendly way. Restraining myself from beating my child takes a lot of mental space and is very hard especially when I am angry, but as the campaign explained, the hard way is going to help me build a healthy relationship with my child and is very necessary for his mental health…"*

Some users in their stories also focused upon their increased understanding about socio-ecological issues like the impact of human activity such as deforestation for mining and industrialization on the environment.

- Economic awareness – Stories also focused upon enhanced awareness and learning about aspects that influenced people's livelihoods, and possibilities to generate income with a view to improve their economic well-being.

In their stories, users narrated the challenges they face in sourcing local jobs, forcing them to remain unemployed, underemployed and in extreme cases migrate out of their villages and towns. Hence, information about local job vacancies as well as employers (that volunteers and MV content team sourced for them on the platform) was something that was particularly relevant to them.

32-year-old, part-time employed MV user from Potka, East Singhbhum reported how listening to the employment news segment helped him remain updated of local job opportunities.

*"…I am employed part-time. When my requirement ends, I become unemployed again. I am very happy that MV has begun providing information on local job opportunities now…I have listened to a number of them and have also shared the relevant ones with my friends. So far, it has turned out to be very helpful as I got to know about two different opportunities – one as a sales man and other as a store operator. I am applying for both the positions... It was very difficult to gather this kind of local information before. Most of the times we didn't get to know if any opportunity existed and at other times, we would get the information so late that by the time we would get to know about a particular job, the candidate would already be chosen."*

Users also valued information about government subsidies in agriculture – such as subsidy on seeds, on high-tech agricultural implements such as drip irrigation systems, and complimentary practical training sessions on cultivation techniques that are not resource intensive, as these helped them increase their yield with minimum resource inputs.

MV took on the role of a financial counsellor as its information on issues like rural banking and financial inclusion services substantiated by fellow users' experiences provided guidance about money matters to the users who heard this content. This led them to have a simplified understanding of financial services and helped them better comprehend the scope of organizing themselves into Self Help Groups for saving and lending small amounts of money. They also reported better awareness of the procedure of saving their money in banking institutions.

- Political awareness – Learning about politico-legal aspects including awareness about governance processes, electoral processes, political parties and their manifestos, was an aspect on which stories were replete. Users attributed to MV their enhanced awareness of various laws such as the Domestic Violence Act, Dowry Prohibition Act, Prohibition of Child Marriage Act, the Right to Information Act, legalities of Corporate Social Responsibility activities for mining regions, and the Land Rights Act. Their interaction over MV further enabled a thorough understanding of different sections of these laws about which they had little or skewed understanding before.

A 38-year-old MV user from Topchanchi, Dhanbad, Jharkhand duly stressed upon the simplification and contextualization of the content of these complex laws in going a long way in helping their communities become better aware of their rights and entitlements, thereby affecting their ability to demand them in future.

*"...Earlier these Acts used to remain confined to the complex law books that people like us living in villages couldn't comprehend. MV has not only simplified them but also put them in an audio format, helping such crucial information reach out to even lesser educated users. In fact, whenever there's an argument in the village, this legal information comes handy in Panchayat meetings for settlement of the issue. As I am able to state the details of the law, people listen to what I have to say and also agree to the settlement. This has positively affected our ability to defend and demand our legal rights."*

Stories revealed that by listening to the content on MV, users were able to add to the information they had regarding their entitlements from the various government programs, more specifically, the rural employment guarantee scheme; scheme for providing Mid-day Meals in schools; subsidized ration distribution scheme; maternal and child health entitlements from various health schemes, and the Right to Education Act. With large number of users belonging to marginalized community groups, this information was especially valuable as it enhanced their ability to demand their entitlements.

Several users in their stories also reflected upon the aspects of political participation such as awareness about the election process, general voter awareness, ideologies of different political parties, reservation in Panchayati Raj elections for women, and the process of filing their candidature for elections. They emphasized how conversations on MV enabled them to better know the candidate and their track record as well as appreciated the potential of MV in being an unbiased platform where different candidates could read their manifestos and pitch themselves to the communities. This not only helped the candidates in reaching out to voters far and wide, but also benefitted many users as most of these users did not have time to go for election rallies and political party meetings, and were able to in an informed and impartial way, decide whom to vote for.

Thus, awareness of these varied political aspects promoted transparency among users about their political ecosystem - their rights, entitlements, and governance, and consequently broadened their perspectives about the political system.

*ii. Introspectional Learning* – Our understanding of introspectional learning draws from Freire's [12] concept of critical conscientization and deals with the praxis of influencing people's thought processes. Introspectional learning is said to occur when individuals exposed to divergent viewpoints understand and critically reflect on them such that they view themselves as the change agents of their own situation. It is not a one-time exercise but a continuous journey of discovery and revelation where they deal with different problems, conflicts, and dilemmas and overcome them to achieve new levels of consciousness. In our analysis, we conceive introspectional learning as a three-step process which includes – (i) critical deliberation and self-reflection; (ii) empathy; and (iii) formation of new ideas, opinions and perspectives. These steps were facilitated by the platform characteristics of *content relevance and relatability*, *editorial credibility* as well as *dialoguing opportunities, which we elaborate now.*

The first step to introspectional learning entails people raising questions about themselves and analyzing their own priorities, their goals as well as those of their immediate and larger groups. It involves critical deliberation and self-reflection to expand their perceptions and world view. Users reported that the content on the platform helped raise critical questions not only about their cultural traditions and practices (e.g. child marriage, women's property inheritance rights) but also about various government policies (e.g. budget, taxation etc.). The dialogue on the platform on these issues led to the users sharing their experiences and opinions. These perspectives were enriched by the joint participation of women and men from varied caste groups and tribes as they brought out the voices of both genders. Several story narrators felt this exchange influenced their views and altered their understanding of the issue and led them to having a more empathetic view of different people and their circumstances. Their stories were full of evidence that portrayed how awareness about the lived experiences of others enabled them to empathize, critically reflect and alter their opinions about various issues/concepts, importantly on issues such as child marriage and violence against women where women's narratives sensitized men and boys against these social evils.

For instance, 15-year-old MV user from Bhatdih Basti, Baghmara, Dhanbad in Jharkhand narrated the process that made him alter his opinion about the practice of 'woman-beating'.

*"I used to think it is okay if women are beaten up for doing something wrong. I wasn't aware it is gharelu hinsa (domestic violence). I became aware of the seriousness of this issue due to the MV campaign on mahila hinsa (violence against women) where I heard first hand experiences of women victims… [which were] very painful. I was also able to relate to people's opinions and comments - it made me think and question things around me – how must my mother, my aunt feel if they were beaten up recklessly, how would I feel about the same? I have understood that when I grow up I am going to fight against such social evils…"*

Thus, users' introspections and reflections due to their engagement with the platform led them to form new thoughts and perspectives. Similarly, many users reported that discussions on MV led them to critically understand and form an opinion about different government policies and current affairs. For instance, the UGCs received on the *Janta ki Report* programme present a mixed bag of experiences about how the decision of the Indian Government to derecognize major currency notes in circulation overnight (the demonetization event in Nov 2016) inconvenienced the rural poor, but was still endorsed by many of them on its promise to rein in black money.

Wignaraja [47] notes that active participation of people, especially the poor can facilitate change through their own reflections which leads to the process of consciousness raising and self-transformation. Freire [12] and Melkote and Steeves [30] emphasize on the vitality of dialogue for the process of reflection to take place. They advocate that particular forms of dialogue (inter-personal and small group strategies) are experiential and enable participants to identify and explore issues that have a meaning for them. This also draws attention to the concept of "multiple truths" [36,40] – knowledge systems and perspectives different to one's own which in turn leads to expanded consciousness, power and liberation.

*iii. Instrumental Learning* – We conceptualize instrumental learning as an action-based learning that results from exposure to information, introspection, rational analyses, discussion and skill-based training. We borrow this concept from Giddens's [14] *radical engagement* and Ilsley's [22] *instrumental/didactic learning* to portray how users' engagement with MV led them to acquire communication/articulation and problem-solving skills.

Many users, especially less educated men and women, in their stories, reported a change in their communication abilities after engaging with MV. They reported learning how to articulate their ideas clearly and communicate effectively with different groups of people. They felt that they were able to better compose their messages and record them on the platform. Some of them also reported feeling competent in being able to draft their message for demanding accountability from government functionaries and departments. These users, in their stories, attributed their ability to overcome their apprehensions and articulate themselves better to the training and constant support provided to them by the MV community volunteers as well as their exposure to different styles and formats of UGC recorded by other users.

A 39-year-old MV user from Ghatshila, East Singhbum in Jharkhand reflected on this aspect in her story.

*"… Being a relatively less educated village woman, I have always been very shy of speaking out publically. However, my regular engagement with MV changed that. Listening to MV, I was able to learn a lot from other contributors in terms of their speaking styles, as different users have their own specific styles. Moreover, based on our local volunteers' advice, I have also begun jotting down my points and practice the content before recording it. This has not only enhanced my public speaking skills but has also improved my ability to articulate my thoughts better…"*

Users also reported their ability to draw learning from experiences discussed on MV and apply them to solve their own problems. In their stories, they shared a range of instances of application of such learning – from the domain of agriculture where they reported learning from other farmers' experiences of how to solve pest problems, to tackling issues of gender-based violence and exploitation. Thus, all four platform characteristics seem to have contributed to this learning.

*iv. Social Learning* – Within the existing frame of users' roles, values and beliefs, we use the concept of social learning [1] as their capacity to negotiate with themselves as well as others to gain confidence in their own abilities to achieve their ambitions [29], and in the process also increase their self-respect, recognition, and reframe their identities within their communities [44]. We classify social learning experienced by the users under the heads of – (i) self-confidence; (ii) social identity; and (iii) network and extended relationships.

- Self-confidence – The belief in one's own abilities, forms an important construct for learning and achievement. Many users in their stories reflected on how their engagement with MV led to an increase in their confidence level, thereby encouraging them to take control of their situation and confidently act with a stronger resolve to make a positive difference in their lives. They reported feeling liberated and validated after being able to raise their voices on the platform to demand accountability as well as express themselves, showcase their talents, share their experiences and relevant information, entirely on their own.

29-year-old Phoolkumari Devi, a volunteer from Ormanjhi, Ranchi detailed this aspect in her story.

*"…Earlier I used to sit at the peripheries of all the meetings as I didn't have the confidence to speak up my mind… but speaking on JMV, engaging with fellow volunteers while attending meetings, attending trainings for recording better content and community mobilization workshops have helped me a lot. I can now speak confidently anywhere, be it Gram Sabha (formal gathering in the village) or Aam Sabha (informal gathering). These trainings have improved my capacity – I can now convene meetings, organize community members, lead protests and interact/work with local officials in resolving/following up the issues of our community members…"*

Many users also reported feeling less self-conscious while contributing content on MV as they felt it was less daunting to record their voices over phone privately than speaking out publicly.

- Social identity – Many users shared instances of evolution of their existing identities to form new identities due to their association with MV. Analysis of stories revealed that by expressing themselves on the platform, many users shared their talents (as poets, singers, etc.) which led their communities to view them differently and thus modified their social identity. MV users working in different domains such as community health workers and social workers shared that the domain specific knowledge they receive listening to MV helps them in doing their job better, and in turn strengthens their credibility and existing identity.

Those users who transitioned from being ordinary community members to MV volunteers specifically mentioned that they began to be regarded as local leaders because of their ability to interact with government officials to resolve the issues of their communities.

Bhasha Sharma, a 38-year-old MV volunteer from Ghatshila, East Singhbhum, in Jharkhand shared this aspect in her story.

*"…Issues such as unavailability of hand pumps, malpractices in the provision of mid-day meals, irregularities in BPL card ration entitlements, etc. arise on a day-to-day basis in our community. In the capacity of a volunteer, I have been trained to follow them up. These follow-ups require me to closely work with our local (government) officials. Due to my position as a MV volunteer and repeat interactions, I have established very good working relations with these officials. I take pride in my ability to fearlessly interact with such powerful people and capacity to resolve our local issues. I think that has earned me the identity of a social activist and local leader in my community as people come to me with their issues and ask for my help in addressing them…"*

Many women volunteers acknowledged the rarity of this experience in their stories, since leadership identities and positions within their communities are usually reserved for men. Additionally, MV volunteers and users (ordinary community members) who regularly contributed content on the platform reported that their association with the platform led them to be identified as "citizen journalists" or media entities due to which they were regularly invited for local events and given press releases to be read / recorded on MV.

▪ Network and extended relationships – According to Granovetter [16], strong ties are observed among close family and friend relationships; while weak ties are present among acquaintances. As there is greater familiarity among the individuals with strong ties, it is likely that they will have access to similar networks which will possess similar information. However, individuals with weak ties move in different circles and hence have access to different information and networks. These information and networks act as bridges and connect otherwise disconnected social groups with new ideas and innovation that can help them in various aspects of their life. Therefore, weak ties form more beneficial networks than strong ties.

Story analysis revealed the emergence of weak ties or secondary Gesellschaft relationships [44] among users due to their interaction with diverse groups of people on/via the platform. These weak ties emerged from – (a) user-user relationships via UGC contributed on the platform; (b) user-volunteer relationships via training, capacity building and other support interaction with volunteers and; (c) user-stakeholder/institution relationships via provision of expert information, advocacy and issue resolution efforts on MV. These weak ties constituted a mix of real-virtual and one-sided or two-way relationships which linked people by offering new opportunities for information exchange and building solidarity, especially in case of user-user relationships. As an example of user-volunteer relationship, users' strong ties with their friends and family (especially for women) influenced their decision to associate with MV (in physical capacity) as volunteers or as club members/users and form weak ties. Many users reported perceiving volunteer clubs as support groups which apart from supporting their community members in varied ways, were also responsible for breaking gender stereotypes. For instance, the clubs with more women volunteers enabled women from conservative household to come out of the confines of their houses and participate in the club meetings as well as promoted their virtual participation on the platform.

24-year-old JMV user Yasmeen Khatoon from Ormanjhi, Ranchi elaborated this point in her story.

*"…My husband never allowed me to go out of the house to participate in public meetings, before. After the formation of women's club in our district my husband encouraged me to participate in club activities. [He was] confident that I'll learn something constructive out of it. Participation on JMV and in club activities has not only helped women like me come out of our conservative households but has also provided us with adequate collective support so that we can stand for ourselves for finding solutions to our problems…"*

Since the users were forming weak and strong ties – a net of relationship and support at different times in different situations emerged, which led to the formation of social capital [35]. Thus, the platform characteristics of multi-level engagement processes; dialoguing opportunities; and editorial credibility enabled social learning among the users.

### 4.3.2 Agency Effects

Due to their engagement with MV, users reported having built greater confidence and felt more in control of the possibilities of them being able to solve their problems, influence their surroundings, and their lives. The agency effects took place at two levels – online / virtually on the platform and offline / beyond the platform and are discussed as follows.

*i. Online Agency Effects* – Users in their stories repeatedly focused on the process of collective validation emerging due to their participation on the platform, which led to the emergence of agency. These effects were found to emanate from the *primary level - platform mediated engagement* characteristic of the platform. In this process, if a user shares their grievance on MV, other users would respond to it by recording similar experiences they had while dealing with that issue in their own context, thereby collectively validating the issue. For instance, if a user recorded on the platform that s/he was denied their entitlement of food grains under the subsidized ration scheme, other users would pitch in with their experiences locally from the same region as well from other districts, in the process validating the existence of the issue at the grassroots. This led to the building of confidence and morale of users to demand accountability from various authorities / institutions and seek recourse to their problems.

In his story, Deepak Kumar Pandey, a 29-year-old JMV user from Topchanchi, Dhanbad recounted an incident where the non-payment of wages under the employment guarantee scheme in Gendnawadih Panchayat was validated on JMV.

*"…Once when Baijnath Mahto (JMV volunteer) motivated a MNREGA labourer to record his complaint about the non-payment of his wages on JMV, many other fellow labourers suffering from the same issue also recorded their woes about non-payment of wages on the JMV platform. They also left a call for listeners to join and support the protest they were organizing in front of the Post Office and the Block Development Office. In addition to the MNREGA Mazdoor Manch (labours' union), local residents joined them in protesting for their rights and many callers from across the state expressed their solidarity with the labourers (virtually) over*

*JMV. Fearing further escalation of this matter, local authorities promptly disbursed their due wages…"*

Stories were replete with examples of issues users faced with the subsidized ration scheme, stoppage of mid-day meals in schools, poor service delivery at the primary health centers etc. that other users followed up in their recordings and led to the validation and collaboration of people's experiences. Story analysis further revealed that by using the platform to repeatedly record voice messages and portray a show of support, users were able to build pressure and demand accountability from the concerned authorities / individuals in powerful positions. This online support created pressure in an overt (direct) and a covert (indirect) manner which led to the users feel greater confidence in their abilities to negotiate different issues. For instance, while using the platform to protest for their contractual appointment to be converted to permanent appointments, para-teachers' overtly blamed the Education Minister of the State and carried out pressure building exercises against the government administration by refusing to carry on with their teaching duties. In other cases, social pressure was also built covertly to inflict collective yet indirect damage to the reputation of corrupt individuals, such as speaking about irregularities in giving out ration entitlements (without name-shaming the person) put pressure on the ration shop owners, and speaking about poor service delivery at primary health centers put pressure on the health workers and authorities. Thus, the platform became another outlet where users could openly discuss their issues and were able to virtually organize themselves to seek accountability.

*ii. Offline Agency Effects* – Stories also emphasized the importance of offline linkages and partnerships developed with different stakeholders at various levels to be vital for the agency effects perceived among the users. These partnerships were developed as a result of the *volunteer mediated engagement* and *institutionally mediated engagement* created through the platform.

▪ Community mobilization: Users in their stories indicated that the offline activities spearheaded by the volunteers, converted the virtual support into actual mobilization of people for specific issues and problems. Apart from engaging with other stakeholders, volunteers also helped organize community members to portray resistance and collective action. Stories were replete with the mention of offline protests led by volunteers on issues demanding better roads, proper sewage disposal and shutting shops in their region that sell alcohol – and how the platform was used intensely for coordination and further mobilization.

In her story, Deepika Devi, a 27-year-old JMV club user from Ormanjhi, Ranchi shared how volunteers networking with officials in government administrative positions have helped resolve their issues.

*"…The real challenges we face are created by small infrastructural issues such as lack of sewage disposal system in our village, lack of proper roads, lack of functional hand-pumps due to which we have to walk kilometers to fetch water. We repeatedly record and discuss about these issues on JMV, but only when club volunteers from our community took initiative of speaking with the BDO (Block Development Officer) and other administrative authorities that we began to see the initiation of work at the grassroots level. The hand-pumps in my village have begun functioning and we have been assured that arrangements for inserting sewer lines around our villages are underway…"*

▪ Accountable and transparent government administration: Only a few government bureaucrats and functionaries agreed to directly liaison with MV. Cases of irregularities in the functioning of government schemes that went beyond the capacity of volunteers to address them at the local level, were forwarded to these higher officials who channeled appropriate action. This was seen with issues relating to the denial of migrant labour registration at the village council level, denial of entitled ration to the public, irregularities in the delivery of mid-day meals, rural employment, rural housing, etc. The officials acknowledged users' issues and responded to them. This action tremendously boosted people's agency and confidence as it demonstrated that their voices were heard and responded to, by senior officials.

32-year-old JMV user Umesh Kumar Turi from Baghmara, Dhanbad, shared how a widow pension case which was reported on JMV was resolved with assistance from the State Labour department.

*"…We raised the issue of the death of a migrant labour on JMV during a campaign, who died while migration (on worksite) due to poor working conditions… JMV interfaced the issue with the Labour department of Jharkhand in Ranchi from where the Labour commissioner personally looked into the issue… commissioner further used the platform to inform all of us to register ourselves in our village councils before migrating out… When we informed them that our village council members haven't received training on the documentation process, the Labour department conducted migrant registration awareness and trainings camps in many blocks and districts including ours…since then they have actively been following up on our migration related grievances/issues that we report…"*

▪ Media convergence: Another aspect contributing to the strengthening of people's agency was the partnership that MV formed with mainstream media. This partnership enabled them to bring attention to issues of the poor, build pressure, and resolve issues that MV users and volunteers were unable to do on their own. For instance, Jharkhand Mobile Vaani (JMV) ran a campaign to collect data on the quality of health services provided at the local health centers across three districts of Jharkhand. The campaign came back with findings that almost 90% of the health centers did not have clean drinking water, more than 50% of posts for doctors were vacant, and doctors were often absent from the clinics. These findings were featured by a leading Hindi regional newspaper. The clout of mass media pushed the authorities to improve the health services. Many users, in their stories, reported improvements in the health facilities in their region upon publication of the news report. Similarly, Farkeshwar Mahto from Topchanchi, Dhanbad reported an incident in his story that portrayed how JMV's convergence with the mainstream media helped in holding the governing institutions accountable and stopping the illegal practices of stone quarrying mafia in his region.

*"…For the past 10-15 years there is a mafia in our region that has illegally acquired a license to blast rocks and sells 10 tractors of rocks everyday. Clearly, the mafia had reached an agreement with the mining department (contractors) and taken over a strip of the land. This piece of land is located near Jamunia river. While heavy blasting polluted the river, the matter began deeply concerning us when the foundations of a newly constructed 3.5 crore bridge (over Jamunia river) began to get affected due to consistent rock blasting… This bridge is the only easy way for us to cross the river. It reached a point that further blasting would have led to the bridge collapsing. We locals organized a protest march, gave a written application to the Deputy Commissioner twice, tried to gather mainstream media support around it, but nothing happened. We then raised this issue on JMV. JMV apart from airing our messages multiple times, also forwarded these recordings to Panchayatnama (mainstream media) that featured an article about this issue. All these efforts reiterated the facts and presented a genuinely verified*

*picture of the situation. The issue received so much coverage that it built a lot of pressure on the authorities. This led to the speeding up of investigation on this matter and within one month the results of the area inspection arrived. Such was the impact that the blasting stopped days before the inspector came for reviewing the matter…"*

- This portrays how the convergence of JMV with mainstream media helped advocate issues faced by the JMV users and held government institutions accountable by building wider media pressure on them.

- Support of civil society organizations (CSOs): As CSOs have different expertise and networks, MV partnered with them to help advocate and resolve users' issues. This led users to find new hope and direction which added to their agency. The most popular instances shared in the stories were when users were able to – (a) directly pose questions to the Union Minister for Human Resource and Development and advocate their views for provision of better quality education in the government schools (via OurSay CSO); (b) advocate for the rights of those who have been forcibly displaced due to mining and other development projects with the policymakers (via IndieVoices) and; (c) have their suggestions and recommendations recorded for the letters to be sent to the Prime Minister's Office (PMO) regarding the poor state of health service delivery at the grassroots level and to advocate for equitable land rights for women (via Oxfam India). Users also shared instances like – grievances related to non-payment of wages under the rural employment guarantee scheme (NREGA) forwarded to civil society movement Gram Swaraj Abhiyan; literacy and education related issues shared with Bharat Gyaan Vigyaan Samiti and Saksharta Vahini; grievances pertaining to violence against women, early marriage and dowry related issues shared with women oriented CSOs such as Oxfam, CREA, Breakthrough, Population Foundation of India and local Mahila Mandals for seeking advice, intervention, and / or resolution.

27-year-old Farooq Ansari from Baghmara, Dhanbad reported that he was able to advocate for issues that he and his community members face, at the national level by interfacing with CSO initiatives via campaigns conducted on JMV.

*"… I heard the programme on JMV where a big minister (Shashi Tharoor, Union Minister for Human Resource and Development) answered my question about improving the English teaching in government schools because without learning English village people like me can't progress in today's modern world. I felt happy that he acknowledged my question as one of important questions of the session and at the same time felt proud because I was able to represent our community level situation which may have some influence on policy level decisions…"*

In the same context, pushing for the inclusion of tribal languages in the mainstream educational discourse, 29-year-old Santhali activist Sunil Soren advocated for the inclusion of Santhali as an optional language in schools.

*"…It is well known that we Adivasis (tribals) are very sensitive towards our cultural roots. Most people within our communities don't send their children to government schools because the children are taught in common (mainstream) languages. I could only do my M.A. (post-graduation) because my parents chose to send me to these schools and passed on the knowledge of Santhali language at home. I feel passionately about this issue and hence discuss this with anyone who lends me an ear. The campaign on JMV provided an interesting opportunity where I could advocate it to the Education Minister. I thus shared my recommendation that Adivasi languages such as Santhali should at least be taught in Schedule V (tribal majority) regions like ours as an optional language. It is feasible because unlike other dialect-based languages, Santhali can be read and written in Olchiki script…"*

While most users perceived advantages, some of them also felt the negative effects of raising their voice on the platform in terms of ridicule, abuse and conflicts with vested interest groups and dominant caste groups. Few users in their stories opened up about the ways in which such powerful people have attempted to intimidate them to prevent them from raising their issues on MV. Their stories revealed that although physical assaults were extremely rare, they regularly received criticism on their comments and encountered full-fledged verbal confrontations. It was also found that threats were not just carried out to intimidate those users who chose to raise their voices but also to silence others who dared to do the same. Many women in their stories have cited this as a factor, in addition to the socio-cultural barriers, that prevents them from actively engaging with the platform. Thus, users perceived several advantages and a few disadvantages owing to these threats. However, the benefits overpowered the threats as they marked small shifts of improvements in their lives.

## 5. MV's Theory of Change

In this section we derive MV's theory of change (ToC) by consolidating our understanding of the three heads of platform processes, platform characteristics and effects discussed above. As seen in figure 1, we conceptualize the platform processes of content generation, community mobilization and building institutional linkages as inputs that lead to the expression of certain platform characteristics as outputs, which according to the users motivate them to participate on MV, and enable them to feel the effects in terms of different kinds of individual learning and collective agency as outputs, thereby representing the change. Explaining the ToC, we now elaborate how platform characteristics led to the emergence of learning and agency effects perceived by the users and how any ICTD platform having such characteristics can bring about change within the communities they serve.

First, the characteristic of *content relevance and relatability* helped contextualize information to make it more understandable and actionable. According to Servaes and Malikhao [40] information becomes knowledge when people are able to interpret it in their own context to derive meanings that suit their lives. Users found the knowledge generated on the platform to be useful in relation to their daily lives which eventually translated into informational learning and formed the basis of other learning and agency effects. Second, the platform characteristic of *editorial credibility* helped users trust the information and experiences shared on the platform, which in turn strengthened the basis of the different kinds of learning. It was also a necessary precursor to agency effects via online processes of the platform as it helped establish collective validity of information shared on the platform.

Third, due to their experiential nature, *dialoguing opportunities* helped generate empathy among fellow users and led to their critical conscientization [12,30]. While the dialoguing opportunities were primarily responsible for the emergence of introspectional learning, processes of consciousness raising and self-transformation, they were also critical for the manifestation of informational and instrumental learning and agency effects.

Finally, the *multi-level engagement processes* of the platform were found to play a key role in manifestation of all the effects due to the range of networks that emerged from the linkages that users developed online (on the platform), and the partnerships that the volunteers and MV developed offline (beyond the platform).

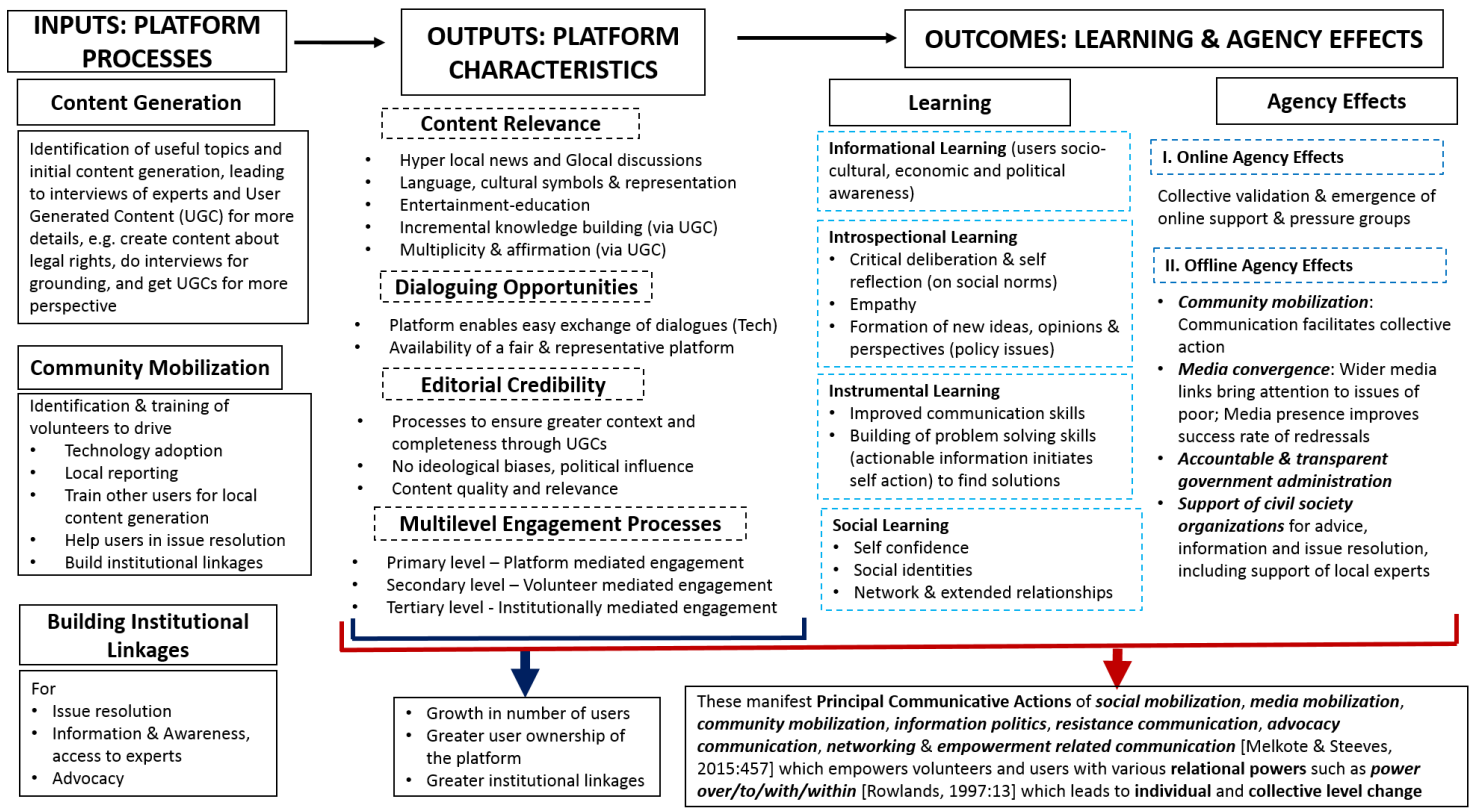

*Figure 1: The Mobile Vaani Theory of Change (ToC)*

Various scholars [4,45,30] have emphasized the importance of networks where access to and inclusion in geographical and/or interest-based networks leads to visibility, control and power, while non-inclusion signifies disempowerment. The multi-level engagement processes thus helped in facilitating a variety of information, experiences, skills and support necessary for users to experience the effects of informational, introspectional, instrumental and social learning as well as online and offline agency effects.

Melkote and Steeves [30] have operationalized Habermas's [18] theory by providing a Principal Communicative Actions (PCA) – Outcomes – Directed change (POD) framework where they outline the various PCA that are manifested when there is an ideal speech situation within the public sphere. With the ideal speech situation enabling individuals to define their own agenda for a dialogue, PCA help individuals to arrive at a consensus and act towards achieving directed change and social justice. Scholars and activists [8,3,5] argue that in today's world, communication technologies form a digital public sphere that enables virtual and more inclusive forms of communication and exchange. MV can thus be conceived as such a digital public sphere, and its editorial policies and inclusive mobilization processes can be considered to reflect the ideal speech situations. As its users belonging to the most economically backward and marginalized communities initiate and participate in the dialoguing processes, it influences their thought processes and leads them to arrive at a consensus, and even act against injustice. This forms the basis for directed change and empowerment [30]. The platform characteristics and effects that user perceive due to their participation on the platform act as kernels for the manifestation of various PCA of – social mobilization, media mobilization, community mobilization, information politics, resistance communication, advocacy communication, networking and empowerment-related communication [30] (as seen in Figure 1) – for triggering change within individuals and communities. These PCA form the bedrock for the emergence of relational powers [37] – *power from within* (strength in abilities of self that inspires and energizes others), *power with* (collective power, power created by group processes), *power to* (generates new possibilities without domination) and *power over* (controlling power) – among MV users. Thus, the platform characteristics led users to engage with the platform and feel certain effects which when observed from the lens of PCA and relational powers, seem to have formed the microcosms of shifts towards empowerment. These microcosms activated at scale could become vehicles of change influencing social development parameters.

## 6. CONCLUSIONS

This paper outlines and validates Theory of Change (ToC) pathways for a mobile phone based, voice-driven platform – Mobile Vaani. The ToC can be generalized and applied to other ICTD initiatives attempting to bring about directed change within their user communities. MV's ToC pathways begin with an input of three *platform processes* of content generation, community mobilization, and building institutional linkages, which lead to the emergence of four *platform characteristics* as outputs. These platform characteristics include content relevance and relatability, editorial credibility, dialoguing opportunities, and multi-level engagement processes, which leads the platform users to experience certain effects or *outcomes*. We have classified these outcomes as different kinds of individual *learning* and collective *agency effects*. We derive the ToC pathways using the concept of

Principal Communicative Actions [30,18] and demonstrate how platform processes manifest into platform characteristics which are at the core of directed social change efforts. We also briefly outline how and when these pathways are rendered broken due to power dynamics arising from patriarchal and caste-based social norms.